\documentclass[conference]{IEEEconf}
\input epsf
\usepackage{graphicx}

\usepackage{cite}
\usepackage{amsmath,amssymb,amsfonts}
\usepackage[hidelinks]{hyperref}
\usepackage{array}
\usepackage{algorithmic}
\usepackage{graphicx}
\usepackage{textcomp}
\usepackage[T1]{fontenc}
\usepackage{tgbonum}
\usepackage{xcolor}
\usepackage{fancyhdr}
\begin{document}

%\title{Submission Format for IPVC-CyberSec21 (Title in 24-point Times font)}
% If the \LARGE is deleted, the title font defaults to  24-point.
% Actually, 
% the \LARGE sets the title at 17 pt, which is close enough to 18-point.
%+++++++++++++++++++++++++++++++++++++++++++
\title{\textbf{\Large A Hierarchical Deep Neural Network for Detecting Lines of Codes with Vulnerabilities\\}}

\author{Arash Mahyari,~\textit{Member, IEEE}\\
	\normalsize Florida Institute For Human and Machine Cognition (IHMC), Pensacola, FL, USA\\
	\normalsize arash.mahyari@ieee.org
}
%+++++++++++++++++++++++++++++++++++++++++++

%\specialpapernotice{(Invited Paper)}

\maketitle
\begin{abstract}
Software vulnerabilities, caused by unintentional flaws in source codes, are the main root cause of cyberattacks. Source code static analysis has been used extensively to detect the unintentional defects, i.e. vulnerabilities, introduced into the source codes by software developers. In this paper, we propose a deep learning approach to detect vulnerabilities from their LLVM IR representations based on the techniques that have been used in natural language processing. The proposed approach uses a hierarchical process to first identify source codes with vulnerabilities, and then it identifies the lines of codes that contribute to the vulnerability within the detected source codes. This proposed two-step approach reduces the false alarm of detecting vulnerable lines. Our extensive experiment on real-world and synthetic codes collected in NVD and SARD shows high accuracy (about 98\%) in detecting source code vulnerabilities.  \footnote{The code repository: \url{https://github.com/arashmahyari/PLP}}.

%\footnote{The code repository: \url{https://github.com/arashmahyari/PLP}}.
\end{abstract}
\IEEEoverridecommandlockouts

% for peerreview papers, inserts a page break and creates the second title.
% Will be ignored for other modes.
\IEEEpeerreviewmaketitle

\fancyhead[RE,LO]{22nd IEEE International Conference on Software
Quality, Reliability, and Security (QRS 2022)}

%\fancyfoot[R]{\thepage}

\section{Introduction}
% no \PARstart

\thispagestyle{fancy}

While preparing source codes, often, there are flaws in the source code that go off the radar of software developers. These flaws lead to vulnerabilities in source codes and open doors for cyber attacks on software, systems, and applications. These vulnerabilities can have disastrous societal and financial consequences \cite{arnold2017after, morrison2018vulnerabilities}. Each year, many vulnerabilities are reported in the Common Vulnerabilities and Exposures (CVE) \cite{cve}.

\begin{figure*}[t]
    \centering
    \includegraphics[width=0.9\textwidth]{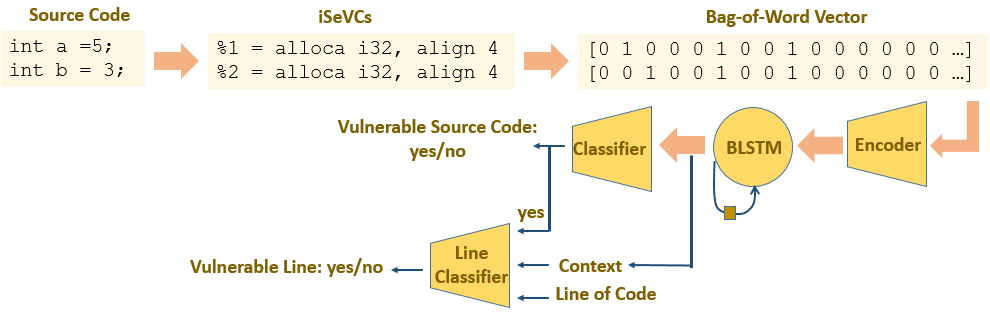}
    \caption{The overall architecture of the proposed vulnerability detection algorithm (PLP I). First, source codes are converted to LLVM IRs, then LLVM IRs are converted to iSeVCs, and then to bag-of-words vector representations. The series of bag-of-words are used as inputs to the encoder, followed by a bidirectional LSTM. The output of the bidirectional LSTM is used to predict whether the whole source code is vulnerable. If so, the hidden states of the bidirectional LSTM is used as the context with each bag-of-words representing an LLVM IR line to predict whether the line is the source of vulnerability.}
    \label{fig:overal}
\end{figure*}

\vspace{2mm}
Source code-based static analysis has been used to detect vulnerabilities. These methods are categorized to several groups, e.g. code similarity-based methods \cite{kim2017vuddy, li2016vulpecker, du2019leopard, wanwarang2020testing}, pattern-based methods \cite{grieco2016toward, yamaguchi2013chucky, yamaguchi2012generalized}, etc. Machine learning-based approaches, including deep learning approaches, belong to the second category \cite{li2021vuldeelocator, li2021sysevr, sedaghatbaf2021automated}. These approaches look for patterns of vulnerabilities in the source codes and are able to generalize these patterns to unseen vulnerabilities. With the success of deep learning approaches in computer vision, a few studies have applied deep learning to source code vulnerability detection \cite{li2021vuldeelocator, li2021sysevr, li2018vuldeepecker}. VulDeePecker \cite{li2018vuldeepecker} was the first approach to incorporate deep learning methods for vulnerability detection. The method only focuses on data dependency among the lines of source code and is not able to achieve fine-grained detection. Syntax-based, Semantics-based, and Vector Representations (SySeVR) \cite{li2021sysevr} was later proposed to overcome the shortcomings of VulDeePecker by introducing semantic-based vulnerability candidates and by representing source codes as vectors. SySeVR is believed to be the first application of deep learning in detecting vulnerabilities. Later, VulDeeLocator was proposed as an improvement to VulDeePecker \cite{li2021vuldeelocator}. VulDeeLocator achieved a finer-grained detection by detecting the pieces of codes causing vulnerabilities. 

\vspace{2mm}

There are a few shortcomings in previous studies and are the motivations of this paper. In previous studies \cite{li2021vuldeelocator}, the Lower
Level Virtual Machine (LLVM) intermediate representations (IRs) lines of codes are treated as sentences of texts that come one after another. However, the line dependencies in source codes are not necessary sequential. For example, in the first two lines of the code, two variables are defined as {\fontfamily{qcr}\selectfont int a; int b;}. Then, in the third line, a value is assigned to the variable defined in the first line {\fontfamily{qcr}\selectfont a=6;}. The third line is related to the first line. This issue is more challenging when it comes to the user-defined functions. Moreover, the existing methods are not able to pinpoint the line of codes contributing to the vulnerability, rather, they can identify the area consisting of several lines \cite{li2021vuldeelocator}. 

\vspace{2mm}

\begin{figure*}[t]
    \centering
    \includegraphics[width=0.6\textwidth]{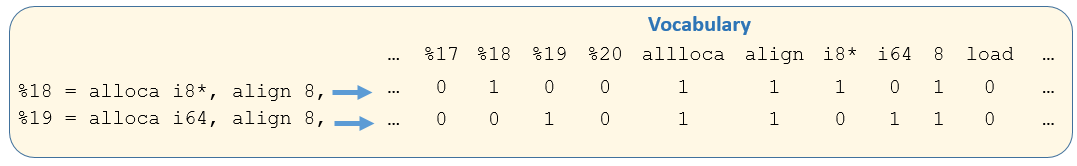}
    \caption{Example of converting LLVM IRs into bag-of-words representation.}
    \label{fig:bag_of_words}
    
\end{figure*}

The contribution of this paper is to develop a Programming Language Processing (PLP) approach, an approach based on natural language processing techniques, to detect vulnerabilities (whole code and lines of codes) based on bag-of-words. We follow the experimental setup in \cite{li2021vuldeelocator} to convert source codes to LLVM IRs before anayzing them for detecting source code vulnerabilities. Then, bag-of-words convert lines of LLVM IR codes into binary vectors and use them as inputs to the deep learning network. The advantage of this approach is that it takes into account the dependencies among the lines of codes. Fig.~\ref{fig:bag_of_words} shows an example of converting two lines of codes into the binary vectors. {\fontfamily{qcr}\selectfont \%1} and {\fontfamily{qcr}\selectfont \%2} appear as two different binary elements in their vector representations\footnote{See Section \ref{sec:data}.}. If there is a line that assigns a value to either of these variables, its vector representation will have a binary one in the location of the variable. After converting LLVM IRs to vector representations, we use a hierarchical approach to detect vulnerabilities. First, a bidirectional long-Short time memory (BLSTM) is used to aggregate the information across all vectors (lines) of a given code and pass the information into a classifier that determines whether the code is vulnerable or not. Once the code is identified as vulnerable, each vector (associated with each line) is passed into another classifier along with its context to classify the line as either good or vulnerable. This hierarchical approach improves the accuracy of the detection and reduces the false alarm because it only looks for vulnerable lines if the code is vulnerable. Fig.~\ref{fig:overal} shows the overall architecture of the proposed approach.

\section{Data}
\label{sec:data}

%\textcolor{red}{need work **********}

We have used the data collected and processed in \cite{li2021vuldeelocator}. The original data are the source codes of C programs from two vulnerability sources: NVD \cite{NVD} and SARD \cite{SARD}. The compatible source codes from these two sources were compiled into LLVM intermediate representations \cite{llvm}. The dataset consists of 14,511 programs (2,182 real-world programs, 12,329 synthetic and academic programs from SARD). The real-world programs are open-source C codes. The synthetic and academic programs are from test cases in SARD. The training dataset includes the real-world vulnerable programs reported prior to 2017, and the test data includes vulnerable codes reported between 2017 to 2019 (unknown vulnerabilities to the training set) \cite{li2021vuldeelocator}.  Two steps are taken to prepare this dataset and convert it to a format that is useful for our proposed method. The deatils of these steps are described in \cite{li2021vuldeelocator}.

\vspace{2mm}

In the first step, source code- and Syntax-based Vulnerability Candidate (sSyVC)s are extracted from source codes. sSyVCs are defined as pieces of code that bear some vulnerability syntax characteristics. In the second step, intermediate code- and Semantics-based Vulnerability Candidate (iSeVC)s are generated from the intermediate codes according to sSyVCs \cite{li2021vuldeelocator}. To extract sSyVCs, the syntax characteristics of known vulnerabilities are represented by abstract syntax trees of the source code. Four types of vulnerability syntax characteristics are used: Library/API Function Call (FC), Array Definition (AD), Pointer Definition (PD), and Arithmetic Expression (AE) \cite{li2021vuldeelocator}.  

\vspace{2mm}

\begin{figure*}[t]
    \centering
    \includegraphics[width=0.8\textwidth]{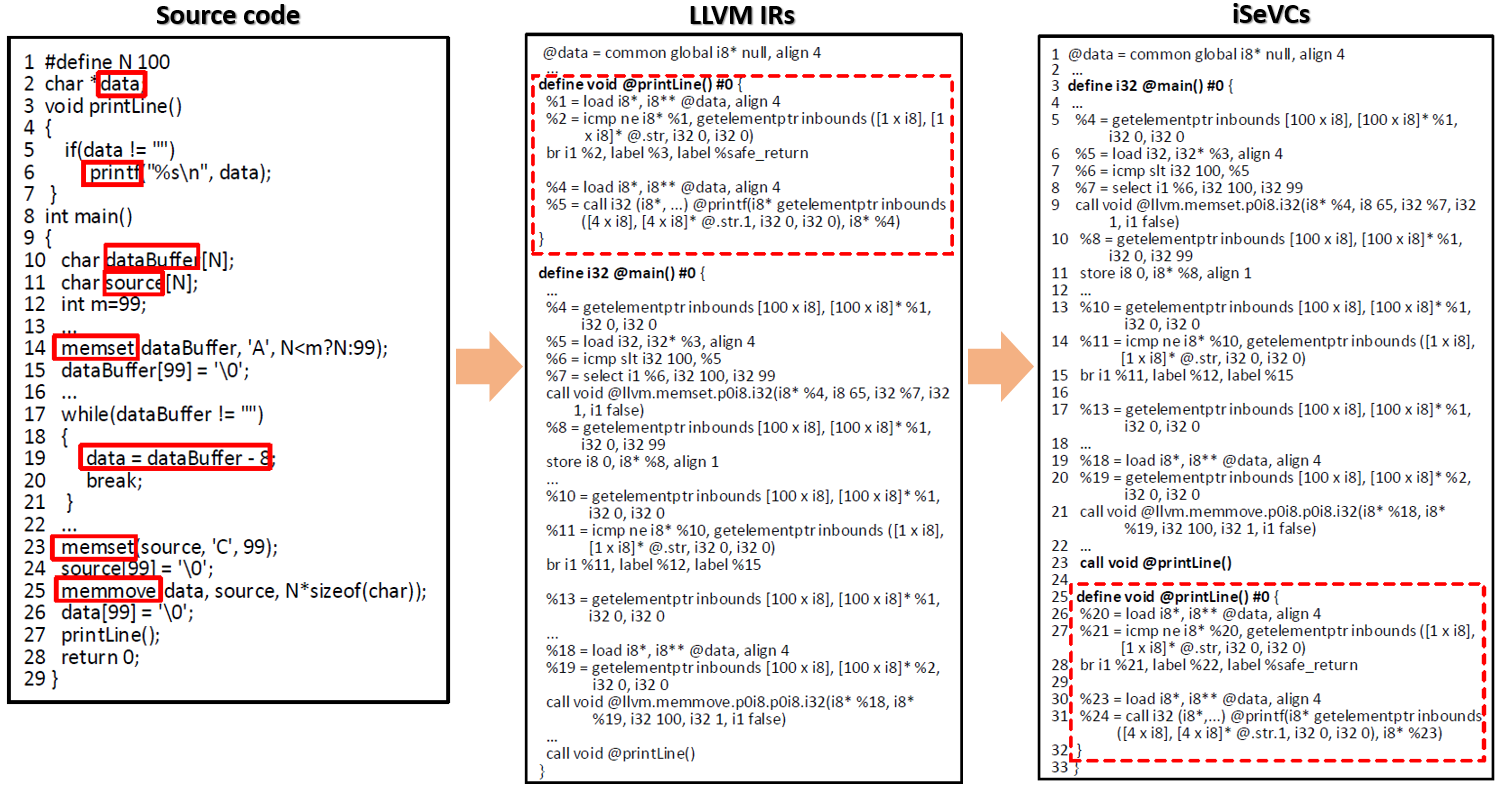}
    \caption{An example source code (on left) is converted to LLVM IRs (in the middle). In iSeVCs, the content of the function "printLine()" (the code inside the red box in the middle) is moved to the line right after calling the function "printLine()" (on the right). Each defined function is immediately followed by its definition. The figure is taken from \cite{li2021vuldeelocator}}
    \label{fig:isevc}
\end{figure*}

In the second step, the Clang compiler is used to generate LLVM bitcode files, link them according to their dependencies, and generate the linked IR files. Given a sSyVC, its dependency graph is generated from its linked IR file and then slice the dependency graph according to the sSyVC (See \cite{li2021vuldeelocator} for more details). Each local variable is converted to a numeric value with a prefix $\%$. For each function $f_\gamma$ called by the function $f_\alpha$, the IR slice of the function $f_\gamma$ is appended to its call in the function $f_\alpha$ (see Fig.~\ref{fig:isevc}). The variable names are adjusted accordingly. The resulted IRs are iSeVCs that will be used for vulnerability detection \footnote{For more information, interested readers may refer to  \cite{li2021vuldeelocator}.}. 

\subsection{Data Preparation}
\label{sec:data-prep}

In this section, we describe the additional preprocessing steps we take to prepare the training and test datasets from LLVM IRs. The performance of sSeVC and iSeVC are compared in \cite{li2021vuldeelocator} and it is shown that iSeVC gives better accuracy. Thus, we use iSeVC provided in \cite{li2021vuldeelocator} in this paper. We use the bag-of-words approach to represent LLVM IRs as vector representations. In the first step, we eliminate all users' defined functions (\verb|"call"| lines immediately followed by \verb|"define"| lines in the processed dataset\footnote{See Section \ref{sec:data}.}). However, we keep the lines within the defined functions. This procedure reduces the number of vocabulary defined by users and makes our algorithm robust toward function names. To speed up the training and evaluation time, we only kept the programs with less than 265 lines of LLVM IRs. This, by no means, reduces the scalability of the algorithm. 

\vspace{2mm}

In the next step, we form a vocabulary of all words separated by a single space, resulting in a vocabulary of size $20,086$ words. This means that there are only $20,086$  different words in all source codes converted to LLVM IR. Then, each line of LLVM IR is converted to a vector of size $20,086$. For each word in the LLVM IR line of code, we place $1$ in the binary vector representing that line of code. The index is the same as the index of the word in the vocabulary. Fig.~\ref{fig:bag_of_words} shows one example of forming bag-of-words for two LLVM IR lines of codes. Note that the advantage of using bag-of-words is that it encodes the order of lines into the vector representation, which ultimately allows the neural network to learn the dependency among the lines of codes. For example, in Fig.~\ref{fig:bag_of_words}, two lines are number 18 and 19 in the source code. These two lines are encoded as a part of the vocabulary into the vector representation.

\vspace{2mm}

If the source code is annotated as a vulnerable code in \cite{li2021vuldeelocator}, we set its label to 1, otherwise, its label is 0. Moreover, the original data in \cite{li2021vuldeelocator} includes the line number of vulnerability. We create a separate binary label for each line of LLVM IR. If the line number matches the source of the vulnerability in the data, we set its label to 1, otherwise it is 0.

\section{Proposed Vulnerability Detector}

The goal of this paper is to detect source code vulnerabilities based on their LLVM IRs. Let ${\bf X}_i=\{ x_i(1), x_i(2), \ldots, x_i(L_i) \}$ represents the $i$th piece of code, where $x_i(t) \in \mathbb{R}^{N \times 1}$ is the bag-of-words representation of the $t$th line, $N$ is the size of the vocabulary, and $L_i$ is the total number of lines in this code. Moreover, let $Y_i \in \{0, 1\}$ represents the label for the $i$th piece of code, where $1$ indicates the code is vulnerable and $0$ means there is no vulnerability in the code. Also, let $\{ y_i(1), y_i(2), \ldots, y_i(L_i) \}$ represent labels for each line of codes, where $y_i(t) \in \{0, 1\}$. Our proposed approach uses a two-step approach to detect whether ${\bf X}_i$ is vulnerable. Then, given a vulnerable code consisting of $L_i$ lines of codes, the proposed approach will identify the vulnerable lines.

\vspace{5mm}

\subsection{Whole-Code Vulnerability Detection}

The proposed network consists of three modules: an encoder, a recurrent neural network (RNN) with Long Short-Time Memory (LSTM) units, and a classifier. The goal of this step is to learn and predict whether the whole source code is vulnerable: $p(Y_i | X_i)$.

\vspace{5mm}

\noindent \textbf{Encoder:} The encoder embeds the binary vector representation of LLVM IRs onto a lower dimensional latent space. The binary vector representation of the LLVM IRs is very sparse and high dimensional. The encoder reduces the dimension of the binary vector representation while making sure that the vulnerable LLVM IRs are discriminated from healthy LLVM IRs in the lower dimension. Because the best weights of the encoder for embeding the vector representation onto a lower dimension are not known, we learn these weights in the end-to-end training.

The encoder consists of two fully connected (FC) layers that embed the $N$-dimensional bag-of-word lines onto a $K$-dimensional vector space:

\begin{equation}
    H_i(t)=ReLU(W_1 \times ReLU(W_0 \times x_i(t))),
\end{equation}

\noindent where $H_i(t) \in \mathbb{R}^{K \times 1}$ is the $t$th line embedding for the $i$th program, $W_0 \in \mathbb{R}^{K \times N}$ is a trainable weight matrix, and \textit{ReLU} is the activation function. %learned from the training data

\vspace{5mm}

\noindent \textbf{LSTM:} The LSTM cell is responsible for learning the sequential pattern of the lines of codes. It is commonly used in NLP to capture the dynamics of language structure and the long term dependency among words in consecutive sentences. We use LSTM modules in the proposed algorithm because the lines of source codes also have long term dependencies. For example, a variable defined in the first line of the code may get a new value in the $25th$ line, thus, the $25th$ line depends on the first line. So, when the source code analyzer gets to this line, it should has a memory to retrieve the effect of the first line. Therefore, we use LSTM modules to capture these long term dependencies.

\begin{table*}[t]
\centering
\caption{The comparison of the proposed method with VulDeeLocator in detecting source code vulnerabilities on the test dataset.}
\label{tab:whole-all-target}
\begin{tabular}{ ccccc } 
\hline
Training set & \multicolumn{2}{c}{Proposed} & \multicolumn{2}{c}{VulDeeLocator \cite{li2021vuldeelocator}} \\
\cline{2-5} 
    & F1\% & Acc.\%
    & F1\% & Acc.\%  \\
\hline
1 &  96.34 & 98.19 & 72.74& 82.43\\
    2 & 98.13 & 99.10 & 72.88& 82.37\\
    3 & 98.07 & 99.07 & 72.67& 82.18\\
    4 & 98.09 & 99.07 & 72.87& 82.36\\
    5 & 98.26 & 99.16 & 72.14& 81.97\\
\hline
%\vspace{2mm}
\end{tabular}
\end{table*}

\begin{table*}[h]
\centering
\caption{The evaluation of the proposed method in detecting vulnerability lines on the test dataset.}
 \label{tab:line}
\begin{tabular}{ ccccc } 
\hline
Training Set & FPR\% & FNR\% & F1\% & Accuracy\% \\
\hline
1 & 22.46 & 0.73 & 85.81 & 97.47\\
    2 & 18.57 & 0.78 & 87.08 & 97.74\\
    3 & 14.87 & 0.83 & 88.35 & 98.00\\
    4 & 20.24 & 0.78 & 86.42 & 97.61\\
    5 & 20.26 & 0.76 & 86.55 & 97.63\\ 
\hline
\end{tabular}
\end{table*}

The LSTM cell can have several layers similar to convolutional and fully connected networks. The LSTM cell takes a sequence of $K$-dimensional vectors (the embedding by the encoder) and calculates the aggregated information across all lines of the code. In this study, we use a bidirectional LSTM (BLSTM) to capture the context of the programming language. Each layer of the BLSTM consists of two LSTM cells (a forward and a backward) that capture information before and after the current line of codes. BLSTM has shown to be more appropriate for capturing the context. Considering the $t^{th}$ line of the code, the forward cell of the BLSTM captures the information from the first line up to the $t^{th}$ line and the backward cell captures the information from the last line all the way back to the $t^{th}$ line of the code, thus capturing what it comes before and after the $t^{th}$ line. The output of the forward LSTM cell at the $t^{th}$ line is $h_F(t)$, and the output of the backward cell at the $t^{th}$ line is $h_B(t)$.

We empirically selected the number of layers of the RNN-BLSTM module. However, the number of layers can be increased for other datasets. The activation function of the forward and backward cells of the BLSTM is \textit{ReLU}. Each LSTM cell (forward and backward) consists of a self-loop, an input gate, a forget gate, and an output gate.

% \begin{align}
%     h_F(t)=ReLU(W_{F_0} \times H_i(t) + W_{F_1} \times h_F(t-1)), \\
%     h_B(t)=ReLU(W_{B_0} \times H_i(t) + W_{B_1} \times h_B(t+1)),
% \end{align}

% \noindent where $W_{F_0}, W_{F_1}, W_{H_0}$ and $W_{H_1}$ are trainable weights, $h_F(t)$ and $h_B(t)$ are the hidden states of the forward and backward networks at time $t$. Thus, $h_i=[h_F(L_i) \hspace{1mm} h_B(1)]^T$, where $T$ is the vector transpose.

\vspace{5mm}

\noindent \textbf{Classifier:} The last output of the forward and backward cells of the BLSTM forms the latent representation of the source code. This latent representation summarizes all information in the source code in single vector representation. The latent representation is classified. The goal of the classifier is to learn and predict the probability of vulnerability, \textit{i.e.} $p(Y_i | h_i)$. The binary classifier consists of two fully connected layers:

\begin{equation}
    {\hat Y}_i(t)=ReLU(W_3 \times ReLU(W_2 \times h_i(t))).
\end{equation}

All three modules are trained end-to-end. The loss function is \textit{cross entropy} with the stochastic gradient descent optimization algorithm \cite{Goodfellow-et-al-2016}.

\subsection{Vulnerable Line Detection}

Given a vulnerable source code, the goal of this module is to learn and predict which lines of the given source code contribute to the vulnerability. This requires a binary classifier that takes each line of code individually and classifies it as vulnerable or not. However, the single line of code may or may not be vulnerable depending on its previous and following lines, \textit{i.e.} the context. Thus, this classifier predicts the vulnerability of the line given its bag-of-word and the context, $p(y_i(t) | x_i(t), C_i(t))$, where $x_i(t)$ is the bag-of-word and $C_i(t)$ is the context. We use the output of the forward cell of the BLSTM at the line $t$ and the output of the backward cell of the BLSTM at the line $t$ to form the context: $C_i(t)=[h_F(t) \hspace{1mm} h_B(t)]^T$. The binary classifier, consisting of two fully connected layers, takes the context $C_i(t)$ and the bag-of-word line of code $x_i(t)$ as the input, and predicts the label $y_i(t)$:

\begin{equation}
    {\hat y}_i(t)=ReLU(W_5 \times ReLU(W_4 \times [h_F(t) \hspace{1mm} h_B(t) \hspace{1mm} x_i(t)]^T)).
\end{equation}

This classifier is trained separately from the previous modules using the \textit{cross entropy} loss with the stochastic gradient descent optimization algorithm. During the evaluation, all modules will be stacked end-to-end.

\section{Experimental Results}
\label{sec:experiment}

In this section, we evaluate the proposed approach to detect source code vulnerabilities and their locations.

\subsection{Detecting Vulnerable Source Code}
\label{sec:exp-whole}

To evaluate the performance of the proposed algorithm to identify vulnerable source codes, we randomly sample from the training dataset and evaluate the trained model on all samples of the target dataset. The number of vulnerable samples in the training dataset is significantly lower than the number of good samples. During the sampling procedure, we sample randomly but equally from each vulnerable and good samples to form the training subset. The trained model is evaluated on all samples of the target dataset. Table.~\ref{tab:whole-all-target} shows the F1 score and accuracy of the proposed method and those of VulDeeLocator \cite{li2021vuldeelocator}. This procedure is repeated 5 times to evaluate the effect of different training subsets. 

For the same training subset, the proposed method achieves higher accuracy and F1 score. This achievement is due to the embedding method used in the proposed approach. The proposed approach is trained end-to-end, which includes the word embedding module (the encoder). Thus, the weights of the word embedding module (the encoder) are adjusted such that the overall network yields higher accuracy. On the other hand, previous methods such as VulDeeLocator \cite{li2021vuldeelocator} uses two step training. In the first step, they train a word2vec model separately from the classifier module, and then a classifier in the second step to detect vulnerable source codes. The higher accuracy of our proposed approach can be attributed to several reasons. Among those, the most prominent one is our proposed way to embed programming words (e.g., for, if) into a latent space. word2vwc approaches provide the best results for natural language processing because the dimension of the vocabulary is extremely large, and one-hot encoding is not possible for all words in the vocabulary. However, programming languages have limited number of vocabulary, which allow us to us one-hot encoding, resulting in a higher accuracy of detection.

\subsection{Detecting Vulnerable Lines of Codes}
\label{sec:exp-whole}

Once we identify vulnerable codes, we use them to detect vulnerable lines in those source codes. Each vulnerable source code has only a few vulnerable lines of codes whereas most lines of codes are good. To balance the number of good and vulnerable lines, in each epoch of training, we use all vulnerable lines and randomly select the same number of good lines of codes. %Similarly, we randomly select the same number of good lines of codes from the target dataset to form the test subset. 
This procedure is repeated five times. Table.~\ref{tab:line} shows accuracy, recall, and precision of the test subset for five iterations (five subsets) separately.

Apparent in the table, the proposed approach achieves a significant higher accuracy and F1 score compared to the previous studies on the same dataset \cite{li2021vuldeelocator}. Previous studies are able to only locate the piece of code containing vulnerabilities. The performance of previous methods are measured by the intersection over union (IoU), approximately $23\%$. Our proposed method achieves the accuracy of $97.6\%$ on average. The advantage of the proposed work is its ability to specify which lines are contributing to the vulnerabilities. Because we know the code is vulnerable (detected in the previous step), the high FPR indicates that the vulnerable line detector stays on the safe side and mark all lines that might be vulnerable. Because our proposed approach is more accurate in pinpointing the line of code, not the approximate area the vulnerability is located, the calculation of IoU is not meaningful for our proposed approach.

\section{Conclusion}

Natural language processing techniques based on deep learning have opened new doors to analyze source codes. These techniques can be leveraged to detect vulnerabilities, recommend next commands, or convert programming languages. In this paper, we proposed the vulnerability detector based on the bag-of-word technique to represent source codes. The source code representations were used to train a bidirectional LSTM to  detect vulnerabilities in source codes. The proposed architecture achieved a significant accuracy thanks to the vocabulary built from LLVM IR source codes. A dedicated study to collect and form a comprehensive vocabulary of LLVM IR commands, and internal calls will facilitate research studies in the future and provide a unique vocabulary that can be used across several source code analysis applications.



% http://www.ctan.org/tex-archive/macros/latex/contrib/supported/IEEEconf/bibtex
%\bibliographystyle{IEEEconf.bst}

%\bibliographystyle{IEEEbib}
%\bibliography{strings_EEG,ref}

%\smallskip
%Note: For the Summary paper submission only, references to the authors own work should be cited as if done by others to enable a double-blind review. {\bfseries Citations must be complete and not redacted, allowing the reviewers to confirm that prior art has been properly identified and acknowledged.}
\end{document}